\documentclass[a4paper,11pt]{article}
\usepackage{pos}
\usepackage{CJK}
\usepackage{amsmath}
\usepackage{dsfont}
\usepackage{graphicx}
\usepackage[linesnumbered,ruled]{algorithm2e}
\DeclareMathOperator{\Tr}{Tr}
\DeclareMathOperator{\Real}{Re}
\DeclareMathOperator*{\argmax}{argmax}

\graphicspath{{./figs}}

\title{Exploring gauge-fixing conditions with gradient-based optimization}

\author[a,b]{William~Detmold}
\author*[c,d]{Gurtej~Kanwar}
\author[a,b]{Yin~Lin}
\author[a,b]{Phiala~E.~Shanahan}
\author[e]{Michael~L.~Wagman}

\affiliation[a]{Center for Theoretical Physics, Massachusetts Institute of Technology, Cambridge, MA 02139, USA}
\affiliation[b]{The NSF AI Institute for Artificial Intelligence and Fundamental Interactions}
\affiliation[c]{Albert Einstein Center, Institute for Theoretical Physics, University of Bern, 3012 Bern, Switzerland}
\affiliation[d]{Higgs Centre for Theoretical Physics, University of Edinburgh, Edinburgh EH9 3FD, UK}
\affiliation[e]{Fermi National Accelerator Laboratory, Batavia, IL 60510, USA}

\abstract{
Lattice gauge fixing is required to compute gauge-variant quantities, for example those used in RI-MOM renormalization schemes or as objects of comparison for model calculations. Recently, gauge-variant quantities have also been found to be more amenable to signal-to-noise optimization using contour deformations. These applications motivate systematic parameterization and exploration of gauge-fixing schemes. This work introduces a differentiable parameterization of gauge fixing which is broad enough to cover Landau gauge, Coulomb gauge, and maximal tree gauges. The adjoint state method allows gradient-based optimization to select gauge-fixing schemes that minimize an arbitrary target loss function. \\

Preprint number: MIT-CTP/5786
}

\FullConference{The 41st International Symposium on Lattice Field Theory (LATTICE2024)\\
 28 July - 3 August 2024\\
Liverpool, UK\\}

\begin{document}

\SetKwComment{Comment}{/* }{ */}
\SetKwRepeat{Do}{do}{while}

\maketitle
	
\section{Introduction}
Gauge fixing is applied in several contexts within lattice field theory calculations, for example to give meaning to gauge-variant observables used in RI-MOM renormalization schemes~\cite{Martinelli:1994ty}, as a computational trick to replace gauge-invariant operators with cheaper gauge-variant operators~\cite{Gao:2023lny}, or as inputs for comparison to phenomenological models \cite{Bonnet:2001uh,Cucchieri:2007rg}. Recently, gauge-variant operators have also been used for contour deformations to reduce statistical noise~\cite{Detmold:2020ncp,Detmold:2021ulb,Lin:2023svo}. In these contexts, the choice of gauge-fixing scheme can affect the efficiency of the calculation, and it may be desirable to systematically explore options for the scheme.

Two kinds of gauge-fixing schemes are commonly used: gauge fixing by functional minimization (e.g.\ Landau and Coulomb gauge) or gauge fixing a maximal tree of links to the identity. Our work makes several contributions in this context. First, we parameterize a continuous family of gauge-fixing schemes that include the former as special cases. Second, we derive the gradients with respect to these parameters of an arbitrary loss function computed from gauge-fixed configurations, which can be used for gradient-based optimization within the family. Finally, we discuss the restriction of this method to a subfamily consisting of maximal trees alone, addressing the discrete nature of this space by introducing a temperature regulator, and demonstrate the effectiveness of this approach in solving two regression problems.

\section{Gauge fixing by functional minimization}
Many standard lattice gauge-fixing schemes are defined by minimizing a functional over possible gauge transformations of a given gauge field. For example, Landau gauge and Coulomb gauge are defined in lattice field theory by respectively minimizing
\begin{equation}
    E_L = -\frac{1}{N_dNV}\sum_{x}\sum_{\mu = 1}^{4} \Real \Tr U^g_\mu(x)
    \quad \text{and} \quad
    ~E_C = -\frac{1}{N_dNV}\sum_{x}\sum_{\mu = 1}^{3} \Real \Tr U^g_i(x),
\end{equation}
where $U^g_\mu(x) = g(x)U_\mu(x)g^\dagger(x+\hat{\mu})$ is the gauge-transformed link, $g(x)$ is the gauge transformation, $V$ is the lattice volume, $N$ is the number of colors, and $N_d$ is the spacetime dimension.
The problem of fixing the gauge field to a particular maximal-tree gauge can be similarly framed into the problem of minimizing the functional
\begin{equation}
    E_{k} = - \frac{1}{N_dNV}\sum_{x,\mu} \Real \Tr k_\mu(x)U^g_\mu(x),
    \label{eq:tree_gcond}
\end{equation}
where $k_\mu(x) \in \{0, 1\}$ is a specified binary indicator field identifying the links $(x,x+\hat{\mu})$ contained in the tree by $k_\mu(x) = 1$.\footnote{In general, \emph{any}
indicator field $k_\mu(x)$ that specifies a subset of links not forming any closed loops corresponds to a tree gauge, in which Eq.~\eqref{eq:tree_gcond} is minimized by fixing all links with $k_\mu(x) = 1$ to the identity.}

All three gauge-fixing schemes can be captured in the general framework of minimizing a gauge functional\footnote{With mild extensions, other gauge fixing schemes, such as maximal Abelian gauges \cite{Kronfeld:1987vd,Bonati:2013bga} requiring minimization of functionals that are multilinear in the gauge field, can also be addressed using the techniques presented here.
}
\begin{align}
    E = - \frac{1}{N_dNV}\sum_{x,\mu} \Real \Tr p_\mu(x)U^g_\mu(x).
    \label{eq:gauge_cond}
\end{align}
The coefficients $p_\mu(x) \in \mathbb{R}$ determine the gauge fixing condition, with the three gauge-fixing schemes discussed above corresponding to specific choices for these coefficients.
Imposing such a gauge-fixing condition corresponds to minimizing Eq.~\eqref{eq:gauge_cond} with respect to $g$. Below, we will study families of gauge-fixing conditions in which the $p_{\mu}(x)$ themselves are parameterized, and these parameters are optimized in order to minimize an independently specified loss function.

We parameterize $g = \exp(i \omega_a(x) T^a)$,
where $a$ is implicitly summed and $T^a$ are the $N^2-1$ generators of $SU(N)$. Because the gauge condition in Eq.~\eqref{eq:gauge_cond} is a smooth functional of $g$, a necessary condition satisfied by the gauge-fixed field $U^{g}_\mu(x)$ is that the derivatives with respect to the parameters of the gauge transformation vanish,
\begin{equation}
    \frac{\partial E}{\partial \omega_a(x)}\bigg|_{U^{g}_\mu(x)} = -\frac{1}{N_d NV}\sum_{\mu}\Real\Tr iT^a\Big(
    p_\mu(x)U^g_\mu(x) 
    -p_\mu(x-\hat{\mu})U^g_\mu(x-\hat{\mu})
    \Big) = 0.
\label{eq:gauge_cond_grad}
\end{equation}
In this work, we do not address the Gribov ambiguity and for the moment assume that each initial gauge field $U_\mu(x)$ can be uniquely mapped to some local minimizer of Eq.~\eqref{eq:gauge_cond}.
In practice, minimization over the gauge transformation parameters $g(x)$ is typically performed iteratively, starting from the initial unfixed gauge field~\cite{Davies:1987vs}, which provides a prescription for such a mapping.

\section{Gradients with respect to the gauge-fixing scheme}
We next define a systematic optimization procedure over the family of gauge-fixing schemes defined by the gauge functional in Eq.~\eqref{eq:gauge_cond}. In different contexts, different features of the gauge-fixing scheme may be important; we choose to capture a general objective via the introduction of a differentiable loss function $\ell[U^g_\mu(x)]$ to be minimized.
Though the loss function is defined directly in terms of the gauge-fixed fields $U^g_\mu(x)$, the gauge-fixed fields carry implicit dependence on the parameters $p_\mu(x)$ defining the gauge-fixing condition in Eq.~\eqref{eq:gauge_cond_grad}. The aim of this section is to make explicit the computation of the gradients $d\ell / d p_\mu(x)$, from which standard gradient descent can be applied to optimize the parameters $p_\mu(x)$.

The standard approach to treat implicit parameter dependence in a differential constraint is to apply the adjoint state method to compute gradients (see~\cite{plessix2006review} for a review).\footnote{This can be contrasted with a brute-force approach in which automatic differentiation is applied to propagate derivatives through the iterations taken to converge to a solution of Eq.~\eqref{eq:gauge_cond_grad}. It has been shown  (for example, Ref.~\cite{hovland2024differentiating} and references within) that such an approach  is generally unstable. The adjoint state method has the added advantage that the iterative solver can be treated as a black box, such that any algorithm can be freely applied.}
Rewriting Eq.~\eqref{eq:gauge_cond_grad} in terms of the gauge transformation parameters $g(x)$, we denote by $g_*(x)$ the parameters that fix the gauge of an input configuration, as defined by
$\frac{\partial E}{\partial \omega_a(x)}\big|_{g(x)=g_*(x)} = 0$.
We can now perform a total derivative with respect to $p_\mu(z)$ to get (all expressions are implicitly evaluated at $g(x) = g_*(x)$)
\begin{equation}
    \sum_{y}\frac{\partial^2 E}{\partial \omega_a(x)\partial \omega_b(y)} 
    \frac{\partial \omega_b(y)}{\partial p_\mu(z)} = -\frac{\partial^2 E}{\partial \omega_a(x)\partial p_\mu(z)}.
\end{equation}
Then the gradient of the loss function is computed as
\begin{align}
\begin{split}
    \frac{d \ell}{dp_\mu(x)} &= \frac{\partial \omega_b(y)}{\partial p_\mu(x)}\frac{\partial \ell}{\partial \omega_b(y)}
    +
    \frac{\partial \ell}{\partial p_\mu(x)}
    \\
    &= -\sum_{y,z}
    \frac{\partial \ell}{\partial \omega_b(y)}
    \bigg(
    \frac{\partial^2 E}{\partial \omega_b(y)\partial \omega_c(z)} 
    \bigg)^{-1}
    \frac{\partial^2 E}{\partial \omega_c(z)\partial p_\mu(x)}
    +
    \frac{\partial \ell}{\partial p_\mu(x)}
    .
\end{split}
\label{eq:adjoint}
\end{align}
The partial derivative $\partial \ell/\partial p_\mu(x)$ encodes the direct dependence of $\ell$ on $p_\mu(x)$ and can be computed explicitly.
The expression in parentheses indicates a matrix inverse over the row/column indices $b,y$ and $c,z$. We can rewrite the expression as
\begin{equation} \label{eq:grad_adj_state}
    \frac{d \ell}{dp_\mu(x)} = -\sum_z\lambda_{c}(z) \frac{\partial^2 E}{\partial \omega_c(z)\partial p_\mu(x)}
    +
    \frac{\partial \ell}{\partial p_\mu(x)}
    ,
\end{equation}
where $\lambda_c(z)$ is the solution of linear equations
\begin{align} \label{eq:adj_state_lambda}
\begin{split}
    \sum_z
    \bigg(
    \frac{\partial^2 E}{\partial \omega_c(z)\partial \omega_b(y)} 
    \bigg)
    \lambda_c(z) &= 
    \frac{\partial \ell}{\partial \omega_b(y)}
    =
    \sum_{x,\mu}\Tr \bigg(\frac{\partial \ell}{\partial U^g_\mu(x)}\frac{\partial U^g_\mu(x)}{\partial \omega_b(y)}\bigg)\\
    &=\sum_{\mu}\Tr \bigg(
    \frac{\partial \ell}{\partial U^g_\mu(y)}
    iT^bU^g_\mu(y)
    - 
    \frac{\partial \ell}{\partial U^g_\mu(y-\hat{\mu})}
    U^g_\mu(y-\hat{\mu})iT^b
    \bigg).
\end{split}
\end{align}
The gradient can thus be evaluated in two steps: first solve for $\lambda_c(z)$ via Eq.~\eqref{eq:adj_state_lambda}, then combine as in Eq.~\eqref{eq:grad_adj_state}.
The relevant derivatives can be computed using the concrete form of the gauge functional from Eq.~\eqref{eq:gauge_cond}, 
\begin{align}
\label{eq:hessian}
   \frac{\partial^2 E}{\partial \omega_a(x)\partial \omega_b(y)}
    &=
    \frac{1}{N_dNV}\sum_{\mu}\Real\Tr 
    T^a
    \Big(
    \delta_{x,y}\big(
    T^bp_\mu(x)U^g_\mu(x) + p_\mu(x-\hat{\mu})U^g_\mu(x-\hat{\mu})T^b
    \big) \\
    &\hspace{20pt} - \delta_{x+\hat{\mu},y}p_\mu(x)U_\mu^g(x)T^b - \delta_{x-\hat{\mu},y}T^bp_\mu(x-\hat{\mu})U_\mu^g(x-\hat{\mu})
    \Big), \nonumber \\
\frac{\partial^2 E}{\partial \omega_a(x)\partial p_\mu(y)}
    &= 
    -\frac{1}{N_dNV}\Real\Tr iT^a\Big(
    \delta_{x,y}U^g_\mu(x) 
    -\delta_{x-\hat{\mu},y}U^g_\mu(x-\hat{\mu})
    \Big).
\end{align}
The Hessian matrix in Eq.~\eqref{eq:hessian} is singular, with a null space of dimension $N^2 - 1$, because the gauge functional in Eq.~\eqref{eq:gauge_cond} is invariant under a global $SU(N)$ transformation of $g(x)$. To break this degeneracy, we can impose the additional constraint that $g(x_0) = M$, where $x_0$ is a point on the lattice and $M \in SU(N)$. The point $x_0$ should then be excluded from the sums above.
By doing so, we remove $N^2-1$ columns and rows from the Hessian matrix such that the resulting matrix, which we will refer to as the reduced Hessian matrix, is invertible. 
In this work, we choose $x_0$ to be the origin of the lattice and $M = \mathbb{I}$. 

\section{Restricting to maximal-tree gauges}

The above method to compute gradients allows one to search over the space of gauge-fixing conditions specified in Eq.~\eqref{eq:gauge_cond} to find the condition that minimizes the gauge-dependent loss function $\ell$. 
Maximal-tree gauges constitute a finite subset of the space of those gauge conditions. 
Motivated by the potential application of using path integral contour deformations to improve the signal-to-noise ratio of observables~\cite{Lin:2023svo}---where improvements to signal-to-noise ratios dependent strongly on the use of certain maximal-tree gauges---we are interested in an algorithm that searches for the minimum of a given loss function within the space of maximal-tree gauges only. 

The discrete nature of the space of possible maximal tree gauges presents a problem to directly applying the algorithm presented in the previous section, as gradients are not well-defined in this subspace. In this work, we choose to circumvent the issue by introducing a temperature parameter $T$ to allow for smooth changes between different trees. The parameters of the gauge-fixing scheme at temperature $T$ are denoted $p_\mu(x; T,v)$ and are defined in terms of a real valued field $v_\mu(x) \in \mathbb{R}$. The temperature dependence is constructed to satisfy 
\begin{equation}
    \lim_{T\rightarrow 0} p_\mu(x; T, v) = k_\mu(x; v),
    \label{eq:p_to_k}
\end{equation}
where $k_\mu(x; v) \in \{0, 1\}$ depends on $v_\mu(x)$ as specified below and is a valid indicator field for a maximal tree. This regularization allows $p_\mu(x; T, v)$ to be optimized at temperatures $T > 0$ where the problem has well-behaved gradients. Several training procedures are possible: for example, one can anneal $T$ towards $0$ to converge towards a fixed maximal tree as training progresses, or a fixed temperature $T > 0$ can be selected for training with the zero-temperature parameters $k_\mu(x;v)=p_\mu(x; 0, v)$ being used only at the time of evaluation.

This training approach relies on a definition of $p_\mu(x; T, v)$ that is computationally tractable, differentiable, and converges as in Eq.~\eqref{eq:p_to_k}. We will call such a function a \emph{soft maximal tree}.
A general method to construct a soft maximal tree is outlined in Ref.~\cite{paulus2020gradient} and  is adapted here.\footnote{Ref.~\cite{paulus2020gradient} uses the equivalent term ``spanning tree'' instead of ``maximal tree''.} In this method, the soft maximal tree $p_\mu(x; T, v)$ is defined as a weighted mean over indicator fields $k_\mu^{(S)}(x)$ for all $S\in\mathcal{S}$, where $\mathcal{S}$ the set of all possible maximal trees. In particular,
\begin{equation} \label{eq:soft-maximal-tree}
\begin{aligned}
    &p_\mu(x; T, v) = \frac{1}{Z(T, v)} \sum_{S \in \mathcal{S}} k^{(S)}_\mu(x) p(S; T, v), \\
\end{aligned}
\end{equation}
where  $p(S; T, v) = \exp\left({\sum_{\mu,x} k^{(S)}_\mu(x) v_\mu(x) / T}\right)$ is the unnormalized probability associated with tree $S$, and $Z(T, v) = \sum_{S \in \mathcal{S}} p(S; T, v)$.
The zero-temperature limit selects the maximal tree $S^*$ that has the maximum total weight in the exponent, i.e.,
\begin{equation}
    \lim_{T \rightarrow 0} p_\mu(x; T, v) = k^{(S^*)}_\mu(x), \quad S^* \equiv \argmax_{S \in \mathcal{S}} \sum_{\mu,x} k^{(S)}_\mu(x) v_\mu(x).
\end{equation}
This limiting maximal tree can be efficiently computed by applying Kruskal's algorithm~\cite{kruskal1956shortest} to the field $v_\mu(x)$.

We are still left with a combinatorially large number of terms in the sum in Eq.~\eqref{eq:soft-maximal-tree}.
Fortunately, the weighted sum of all maximal trees can be computed in polynomial time using (the weighted version of) Kirchhoff’s Matrix-Tree Theorem. The theorem applied to a graph consisting of lattice links weighted by $w_\mu(x; T) = e^{v_\mu(x)/T}$ states that
\begin{equation} \label{eq:kirchoff-theorem}
    \det \overline{L} = \sum_{S \in \mathcal{S}} \prod_{(x,\mu) \in S} w_\mu(x; T) = \sum_{S \in \mathcal{S}} \prod_{x,\mu} e^{k_\mu^{(S)}(x) v_\mu(x)/T} = Z(T,v),
\end{equation}
where $\overline{L}$ denotes the reduced graph Laplacian. The reduced graph Laplacian is obtained by removing $i$th column and row from the graph Laplacian,\footnote{The result is insensitive to the choice of $i$.} which is defined by the matrix elements
\begin{equation}
L_{x,y} = \sum_{\mu} \delta_{x,y}\Big(
    w_\mu(x; T)
    + w_\mu(x-\hat{\mu}; T)
    \Big)
    - \delta_{x+\hat{\mu}, y} w_\mu(x; T)
    - \delta_{x-\hat{\mu}, y} w_\mu(x-\hat{\mu}; T).
\label{eq:lap_1}
\end{equation}
Finally, we can use Eq.~\eqref{eq:kirchoff-theorem} to compute $p_\mu(x; T, v)$ via a derivative,
\begin{equation} \label{eq:p_def}
    p_\mu(x; T,v) =
\frac{\partial \ln \det \overline{L}}{\partial \ln w_\mu(y; T) } =
    T \frac{\partial \ln \det \overline{L}}{\partial v_\mu(x)}.
\end{equation}

In Ref.~\cite{paulus2020gradient}, it was suggested to use automatic differentiation to compute Eq.~\eqref{eq:p_def}.
We can also use the trace identity for the derivative to compute
$p_\mu(x; T,v) = T \Tr 
[\overline{L}^{-1} \partial \overline{L} / \partial v_\mu(x; T) ]$,
where $\partial \overline{L}/\partial v_\mu(x; T)$ can be computed analytically.
Both of these approaches have  computational complexity similar to that of computing the gradient of a fermionic determinant,  scaling as the cube of the number of lattice sites. This direct approach was used for the numerical studies below, but we note that using a stochastic resolution of the determinant---such as the introduction of fields analogous to pseudofermions---may be a useful method to avoid the poor scaling inherent in the calculation of $p_\mu(x; T, v)$ in the future.

\section{Summary of the algorithm}

Combining the details of the previous sections, the key component of minimizing a given loss function $\ell$ via gradient descent is applying the adjoint state method in Eq.~\eqref{eq:adjoint}. For optimizing specifically over the discrete space of maximal tree gauges, we have introduced a temperature to regularize the discreteness of the space, resulting in additional steps in the optimization procedure. The details of this approach are summarized in Alg.~\ref{alg:tree}.
An arbitrary temperature schedule can in principle be used in this algorithm.
For the loss functions $\ell[U^g, p]$ considered in the numerical results below, the parameters $p_\mu(x; T,v)$ converge towards a single maximal tree even while training at a fixed non-zero temperature. For general loss functions, however, it is likely necessary to anneal the temperature $T$ towards zero over time.

\begin{algorithm} 
  \SetNoFillComment
  \caption{Procedure to optimize over maximal-tree gauge-fixing conditions, given a loss function $\ell$, gauge configurations $U_\mu(x)$, and a temperature schedule.}
  \label{alg:tree}
$v_\mu(x) \gets$ initialize the real-valued parameters for optimization 
  \;
  $T \gets$ initialize the temperature
  \;
\BlankLine
  \While{Training}{
    \tcc{Compute the gauge-fixing parameters}
    $\overline{L}[T, v] \gets$ compute with Eq.~\eqref{eq:lap_1} \;
    $p_\mu(x; T, v) \gets$ compute Eq.~\eqref{eq:p_def} by differentiating $\ln \det \overline{L}[T, v]$
\;
    \BlankLine
    $U^g_\mu(x) \gets$ gauge fix by minimizing Eq.~\eqref{eq:gauge_cond}  \label{lst:fix-learned-gauge}
    \;
    $\ell[U^g, p]\gets$ compute the loss function
    \;
    \BlankLine
    \BlankLine
    \tcc{Backpropagation}
    $\frac{\partial \ell}{\partial p_\mu(x)} \gets$ compute with Eq.~\eqref{eq:adjoint}
    \;
    $\frac{\partial \ell}{\partial v_\mu(x)} \gets$ compute using $\frac{\partial \ell}{\partial p_\mu(x)}$ and automatic differentiation
    \;
    $v_\mu(x) \gets$ update with gradient descent
    \;
    $T \gets$ update according to temperature schedule
  }
  \BlankLine
$k_\mu(x;v) \gets$ apply Kruskal's algorithm to $v_\mu(x)$ 
\end{algorithm}

\section{Results for a regression problem}
To demonstrate the effectiveness of Alg.~\ref{alg:tree}, we present the results on a family of regression tasks, defined by optimizing over maximal tree gauges to best reproduce gauge configurations fixed to a given scheme. In each regression task, a target maximal tree $S$ is specified by a fixed indicator field $k^{S}_\mu(x)$; results are presented below for $S$ chosen to be either an axial gauge or a specific randomly generated maximal tree obtained by applying Kruskal's algorithm to a uniform random weight field.
All numerical results are based on an ensemble of $9000$ $SU(2)$ configurations, generated using a Wilson gauge action with coupling $\beta = 4.2$ on a $16 \times 16$ lattice.
For each target $S$, the training data $U^{(S)}_\mu(x)$ are the result of fixing the ensemble to the maximal-tree gauge associated with tree $S$. The target loss function is the mean squared error,
\begin{equation}
    \ell[U^{g}] = \frac{1}{N_dN^2V} \sum_{x,\mu} \Tr |U^{g}_\mu(x) - U^{(S)}_\mu(x)|^2.
\end{equation}
Note that the information $k_\mu^{(S)}(x)$ specifying this tree is \emph{not} explicitly part of the loss function, i.e., this must be inferred by the optimization procedure given only the values $U_\mu^{(S)}(x)$ of the ensemble in the fixed gauge. This is representative of the structure of typical loss functions that might occur in learned gauge-fixing tasks, where the behavior of the gauge-fixed links determines the quality of the gauge-fixing scheme, rather than the geometry of the gauge-fixing tree itself (i.e., the indicator field).

Training is performed via stochastic gradient descent, as specified in Alg.~\ref{alg:tree}, starting from the initialization $v_\mu(x) = 0$. This initial condition results in a constant $p_\mu(x; T, v)$, such that the initial gauge is Landau gauge.
The temperature is fixed to the constant value $T = 1$ throughout training. We use the Adam optimizer \cite{kingma2017adammethodstochasticoptimization} to update the parameters at each iteration, with learning rate fixed to $10^{-2}$. Gradients are estimated using mini-batches of size $32$, and a total of 300 updates applied in each case. In each iteration, the batch of gauge fields are fixed (Alg.~\ref{alg:tree}, line \ref{lst:fix-learned-gauge}) according to the current parameters $p(x; T, v)$ using iterative minimization
until a target tolerance of $10^{-12}$ is reached~\cite{Davies:1987vs}.

\begin{figure}
    \centering
    \includegraphics[width=0.45\textwidth]{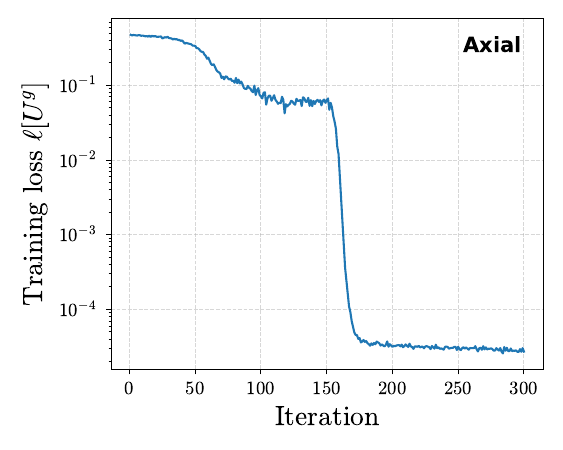}
    \includegraphics[width=0.45\textwidth]{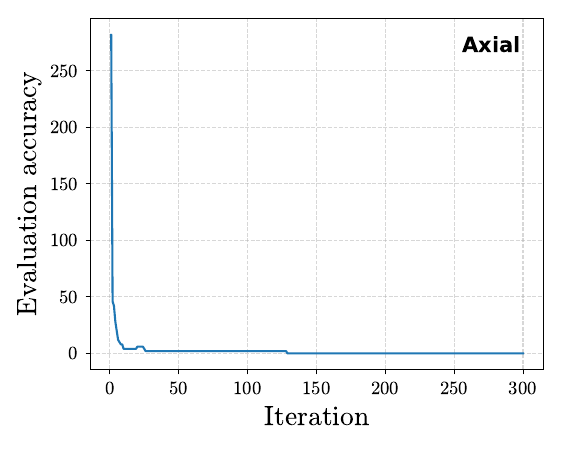} \\
    \includegraphics[width=0.45\textwidth]{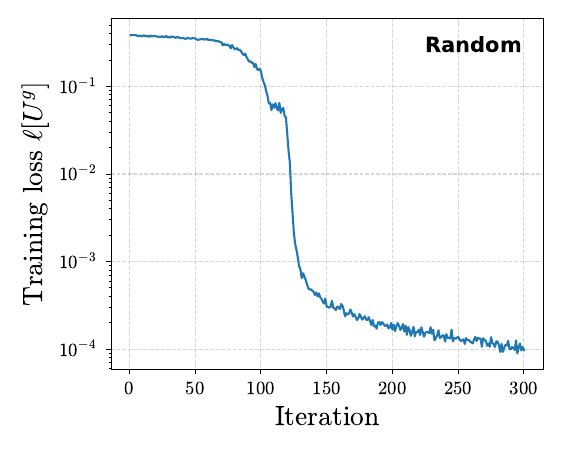}
    \includegraphics[width=0.45\textwidth]{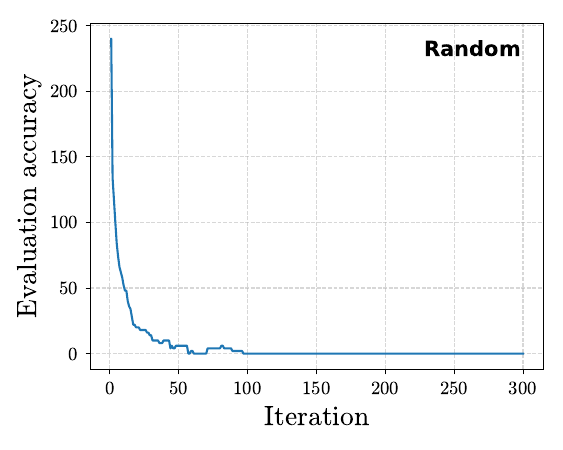}
    \caption{Training history for an axial (above) and randomized (below) target tree.
    The left plots show the training loss as a function of training updates, while the right plots show the accuracy of the learned parameters, quantified as 
    $\sum_{x,\mu} |k^{(S)}_\mu(x) - p(x; 0, v)|$.}
\label{fig:random_tree_line}
\end{figure}

\begin{figure}
    \centering
    \raisebox{-.45\height}
    {\includegraphics[width=0.25\textwidth]{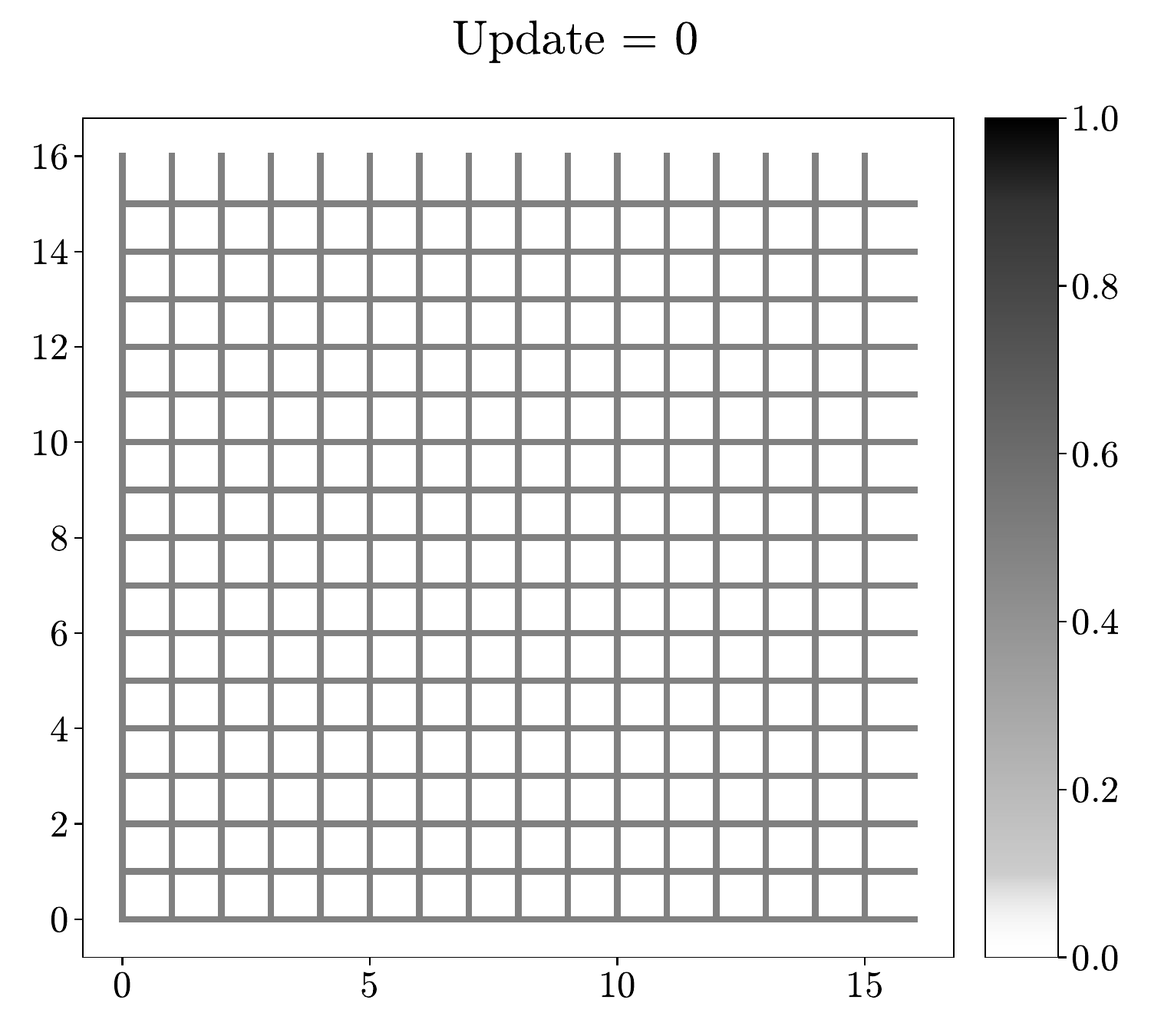}}
    $\to$
\raisebox{-.45\height}
    {\includegraphics[width=0.25\textwidth]{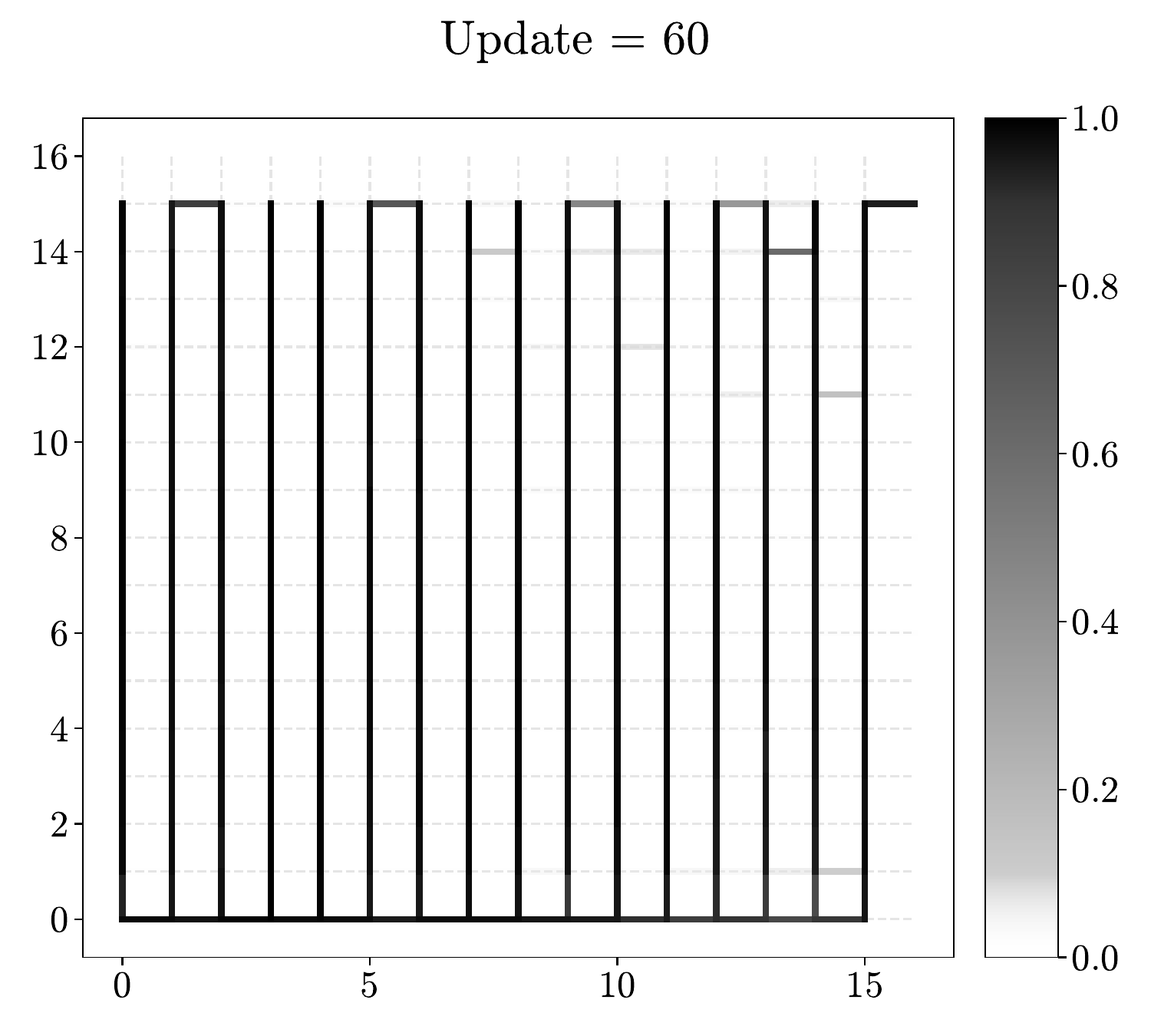}}
    $\to$
\raisebox{-.45\height}
    {\includegraphics[width=0.25\textwidth]{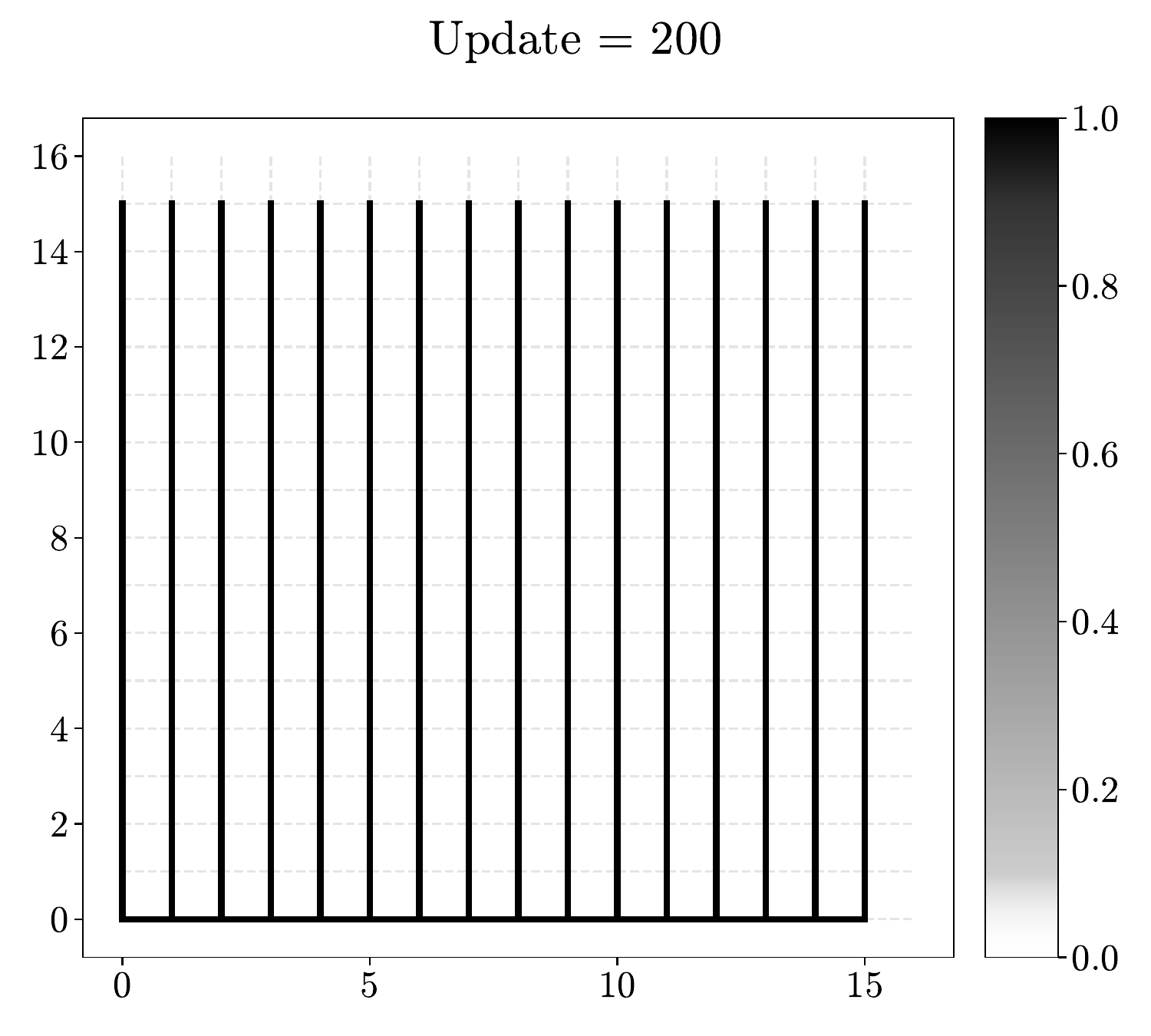}}\\

    \raisebox{-.45\height}{\includegraphics[width=0.25\textwidth]{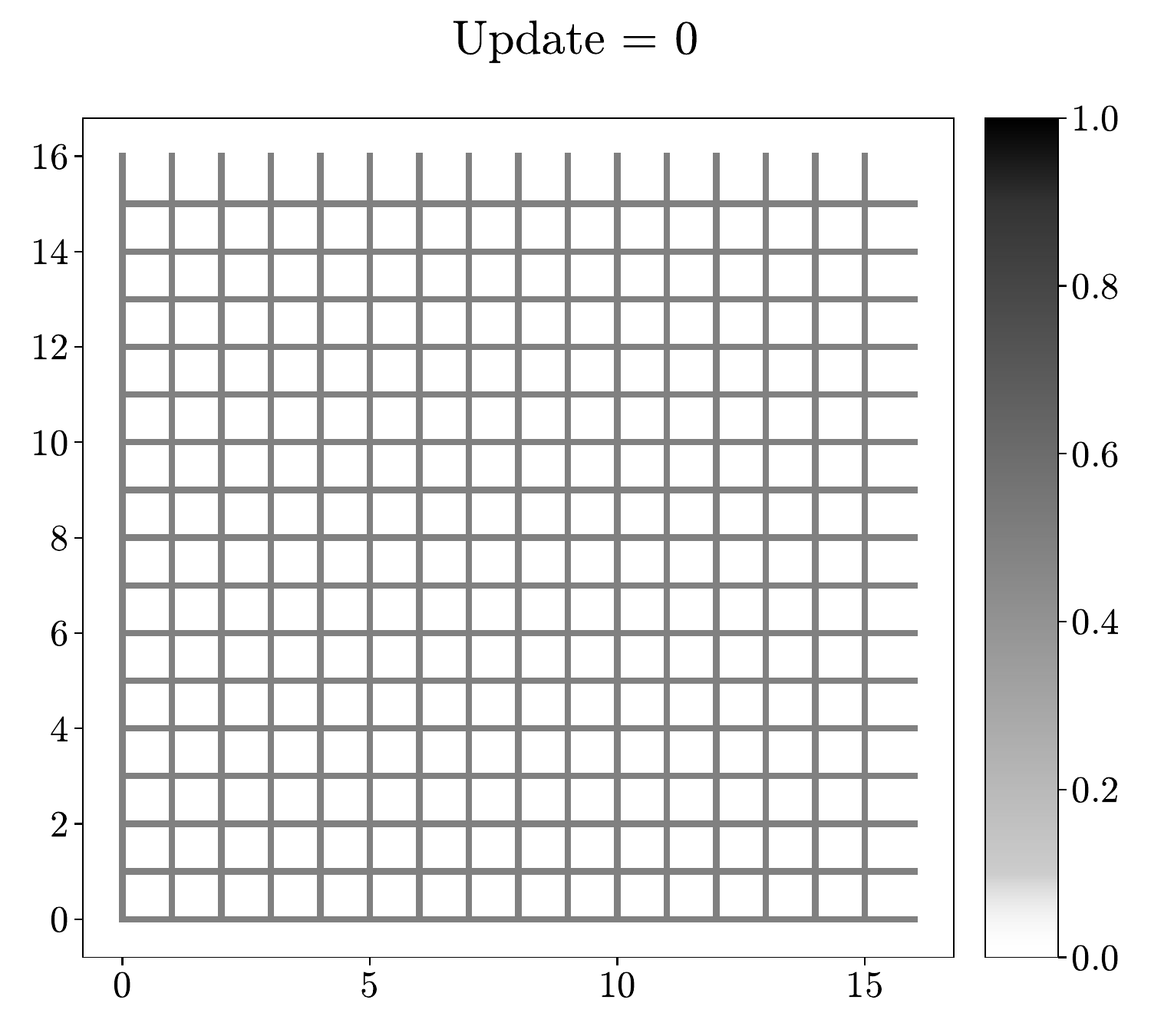}}
    $\to$
\raisebox{-.45\height}{\includegraphics[width=0.25\textwidth]{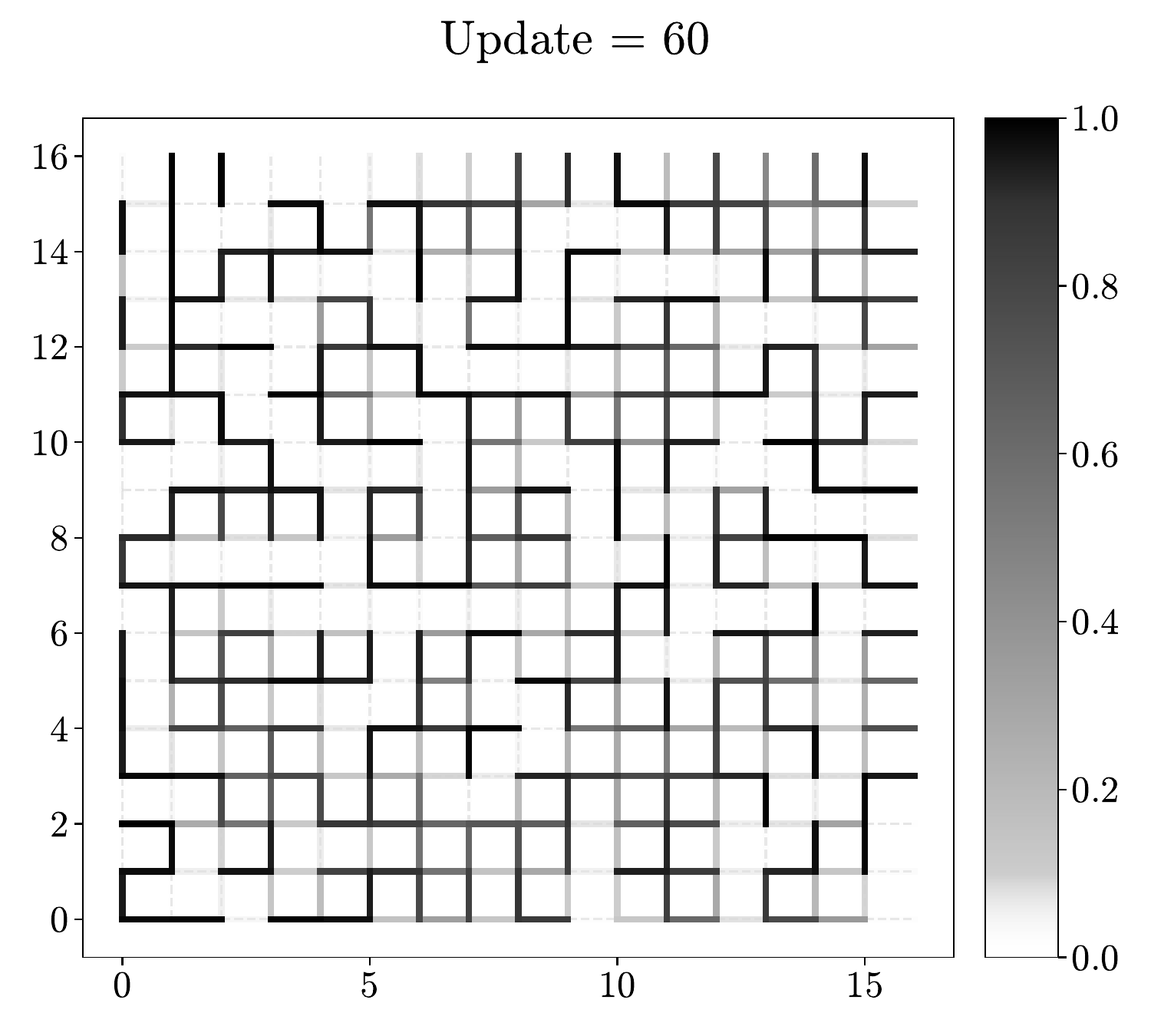}}
$\to$
    \raisebox{-.45\height}{\includegraphics[width=0.25\textwidth]{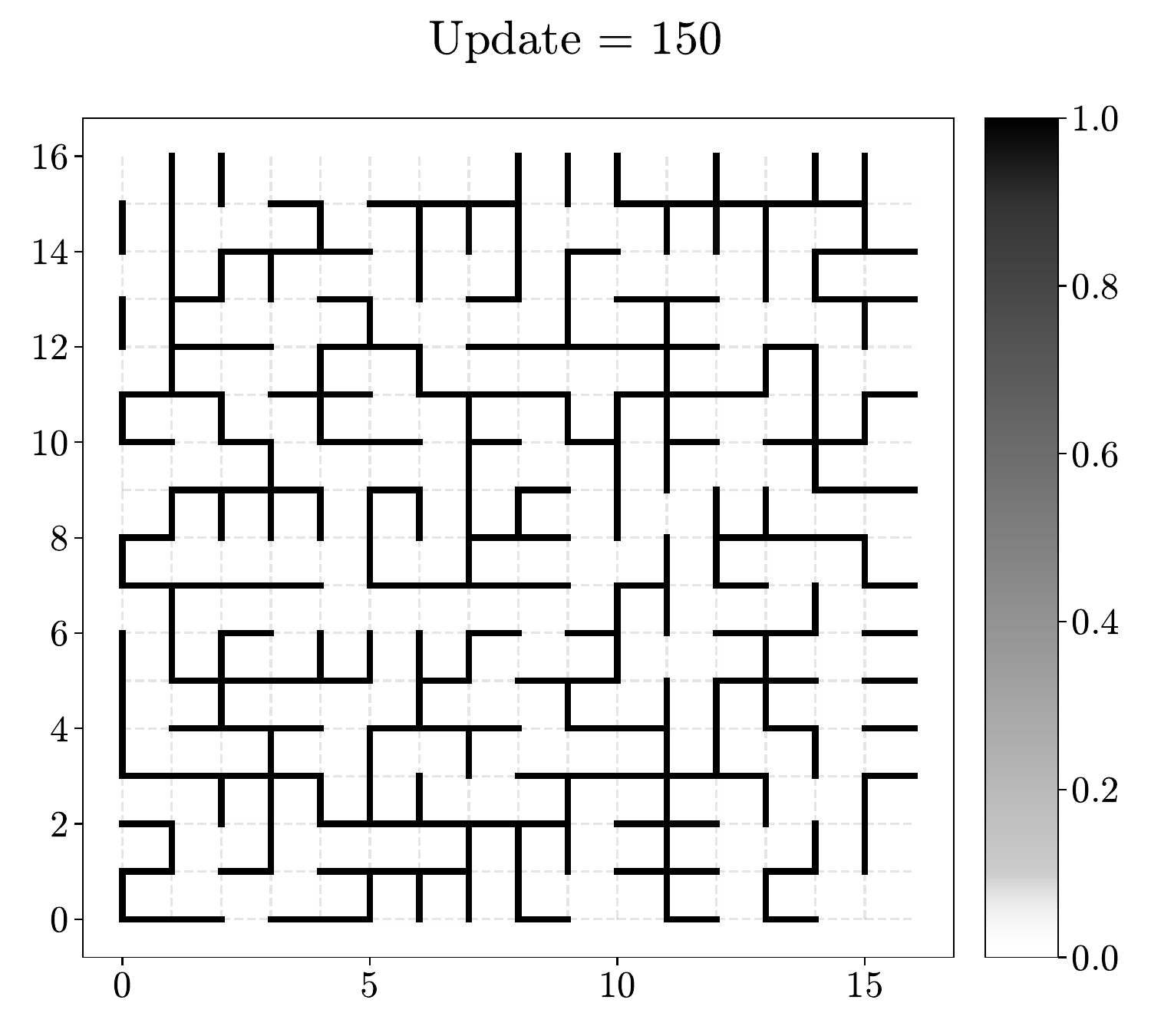}}
    \caption{Snapshots of $p_\mu(x;T,v)$ at various points in the training history for an axial (above) and randomized (below) target tree, corresponding to the respective plots in Fig.~\ref{fig:random_tree_line}.}
    \label{fig:random_tree_p}
\end{figure}

We show results for regression on the two choices of target maximal trees in Fig.~\ref{fig:random_tree_line}.
Figure \ref{fig:random_tree_p} shows snapshots of $p_\mu(x;T,v)$ at various points in the training history for both target trees.
For both targets, the algorithm is able to reproduce the target exactly in less than $150$ updates to $v_\mu(x)$, indicating that the procedure in Alg.~\ref{alg:tree} successfully propagates useful gradient information from the loss function to the soft maximal tree parameterization.
These results verify the consistency and utility of the gradient propagation scheme proposed in this work, suggesting it may be of use in contexts where a systematic exploration of maximal tree gauges is desirable.

\bibliographystyle{utphys}
\bibliography{bib}
\end{document}